\documentclass[MSNbibl,nameyear,dvips]{arxstspdf}
\usepackage{flushend}
\usepackage{stfloats}
\usepackage{graphicx}

\volume{29}
\issue{3}
\pubyear{2014}
\firstpage{425}
\lastpage{438}
\doi{10.1214/13-STS464} 

\makeatletter
\setattribute{abstract}{width}{342pt}
\makeatother

\begin{document}
\begin{frontmatter}

\title{A Conversation with Howell Tong}
\runtitle{A Conversation with Howell Tong}

\begin{aug}
\author[a]{\fnms{Kung-Sik}~\snm{Chan}\corref{}\ead[label=e1]{kung-sik-chan@uiowa.edu}}
\and
\author[b]{\fnms{Qiwei}~\snm{Yao}\ead[label=e2]{q.yao@lse.ac.uk}}
\runauthor{K.-S. Chan and Q. Yao}

\affiliation{University of Iowa and London School of Economics and
Political Science}

\address[a]{Kung-Sik Chan is Professor of Statistics,
Department of Statistics \& Actuarial Science, University of Iowa, Iowa
City, Iowa 52245, USA \printead{e1}.\vspace*{6pt}}
\address[b]{Qiwei Yao is Professor of Statistics, Department of Statistics,
London School of Economics and Political Science, United Kingdom
\printead{e2}.\vspace*{9pt}}
\end{aug}

%
\begin{abstract}
Howell Tong has been an Emeritus Professor of Statistics at the London
School of Economics since October 1, 2009.
He was appointed to a lectureship at the University of
Manchester Institute of Science and Technology shortly after he started
his Master program in 1968. He
received his Ph.D. in 1972 under the supervision of Maurice Priestley,
thus making him an academic grandson of Maurice Bartlett. He stayed at
UMIST until 1982, when he took up the Founding
Chair of Statistics at the Chinese University of Hong Kong. In 1986, he
returned to the UK, as the
first Chinese to hold a Chair of Statistics in the history of the UK,
by accepting the Chair at the
University of Kent at Canterbury. He stayed there until 1997, when he
went to the University of
Hong Kong, first as a Distinguished Visiting Professor, and then as the
Chair Professor of Statistics. At the University of Hong Kong, he served
as a Pro-Vice-Chancellor and was the Founding Dean of the Graduate
School. He
was appointed to his Chair at the London School of Economics in 1999.
He is a pioneer in the field of nonlinear time series analysis and has
been a scientific leader both in Hong Kong and in the UK. His work on
threshold models has had lasting influence both on theory and
applications. He has drawn important connections between time series
and deterministic dynamical systems, linking statistics with chaos
theory, and the models he has developed have found significant
applications in fields as diverse as economics, epidemiology and
ecology. He has made novel contributions to nonparametric and
semi-parametric statistics, especially in model selection and dimension
reduction for time series data.
He has written four
books (one with Kung-Sik Chan and another with Bing Cheng) and over 162
papers (sometimes with collaborators) in Statistics, Ecology,
Actuarial Science, Control Engineering, Reliability, Meteorology, Water
Engineering, Engineering
Mathematics and Mathematical Education.
His 1990 monograph \emph{Non-linear Time Series Analysis---A Dynamical
System Approach} is a classic.
He is a Foreign Member of the Norwegian Academy of
Science and Letters, an elected member of the ISI, a Fellow of IMS and
an Honorary Fellow of the Institute
of Actuaries (UK). He won a Chinese National Natural Science Prize
(Class II) in 2000 and the Royal Statistical Society
awarded him the Guy Medal in Silver in 2007.\looseness=-1

The following conversation is partly based on an interview that took place
in the Hong Kong University of Science and Technology in July 2013.
\end{abstract}

\end{frontmatter}
\newpage\eject

 \textbf{QY:} You were supervised by Maurice Priestley for
your doctorate. What was your thesis on?

 \textbf{HT:} My doctoral thesis was entitled ``Some problems
in the spectral analysis of bivariate nonstationary stochastic
processes.'' It was an extension of Maurice Priestley's evolutionary
spectral analysis, which he proposed in 1965, from the univariate case
to the bivariate case, including both the open-loop and close-loop
systems. The contents of the thesis formed the basis of a joint paper
which Maurice and I read to the Royal Statistical Society in 1972. I
can still remember the occasion well, as it was my first taste of
academic subtlety in Britain.

\begin{figure}

\includegraphics{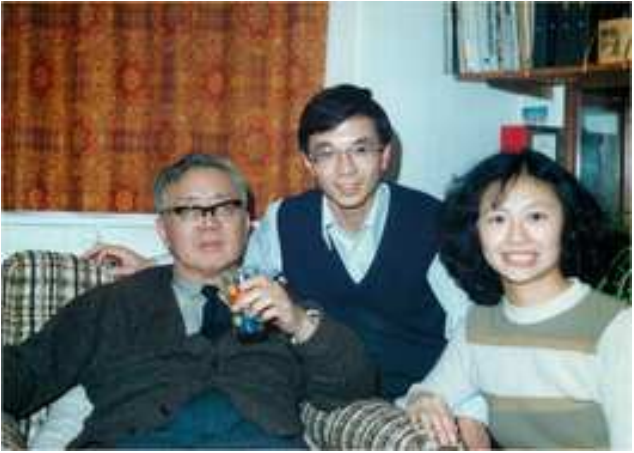}

  \caption{Howell with his childhood hero, Professor Loo Keng HUA, and Mary Tong, at Tong's home in Poynton, Cheshire, UK, 1979.}\label{fig1}
\end{figure}

I must tell you that a British statistician can do a clean demolition
job at an RSS discussion meeting, without even showing his hammer. (I
hope you will forgive me for being gender blind when I speak.) It has
been said that one has to be courageous or foolhardy to read a paper to
the RSS. I have learnt a lot besides the demolition skill since then,
by attending RSS discussion meetings in London. The frankness of views
is very helpful, as much for the readers as for the authors, because it
enables everybody to have a more critical assessment of the strengths
and weaknesses of the presented work. Of course, there will always be
cases of premature euphoria as well as cases of misplaced cold\vadjust{\goodbreak}
shoulder. Despite its imperfection, I do not think that I am alone in
saying that the forum remains the best in the statistical world. In
many ways, it has made the RSS unique.

Returning to my doctoral thesis, I think much of it is now out of date
and mostly of little practical significance. I am especially
disappointed with the fact that the evolutionary coherency spectrum for
nonstationary time series turns out to be time invariant. However,
there is perhaps a curious little result in the thesis which you might
find interesting. It concerns the function $\exp\{i(k\omega t + \omega
_0 t^2)\}$, ${\omega}_0$ being a fixed constant. I~showed that this
frequency-modulated wave admits no generalized frequency in Priestley's
sense. In fact, I am inclined to take the view that for
frequency-modulated waves the wavelet approach is more natural. In the
1990s, Bing Cheng and I developed a wavelet representation for a
general stochastic process.

For the modelling of nonstationary time series, I~think that the
piecewise stationary approach introduced by Tohru Ozaki and myself in
1975 is a very practical one. Specifically, as each new ``short'' block
of data arrives, we check if the AR model fitted to the latest block
needs to be changed. If it does, then a new AR model is the latest
state of the system, otherwise the previous state stays. This approach
is ideally suited for real-time implementation. I understand that
Professor Genshiro Kitagawa and his marine engineering colleagues have
built many successful auto-pilots for boats based on this approach,
under the guidance of the late Professor Akaike.

 \textbf{QY:} Can you tell us something about the early part
of your career in higher education?

 \textbf{HT:} My first job in higher education was a
lectureship at the then Northern Polytechnic, London, UK, in 1967.
Remember I had only a B.Sc. degree! I~took the job for two reasons: (1)
To help my father financially because my mother had just joined us in
London from Hong Kong, having waited for seven long years; (2)~I~lost
my passion for Algebra.

When I graduated from the University of Manchester Institute of Science
and Technology (now merged with the University of Manchester) in 1966,
I was very keen on Algebra. So I went to Queen Mary College of the
University of London on a postgraduate studentship funded by the UK
Science and Engineering Research Council. The general expectation was
to do a Ph.D. in Algebra.

At that time, QMC was the hot house of Algebra in the UK, under the
inspiring leadership of Professor Kurt Hirsch. He came to the UK to
escape from Hitler's Germany, like many of his contemporaries including
Bernard Neumann, Hanna Neumann, Paul M. Cohn and others. He was my
mentor. I remember \mbox{attending} courses on Homological Algebra, Group
Representation Theory and others. I even attended seminars given by
Saunders MacLane and other leading algebraists. One of the first things
that Professor Hirsch asked me was ``Have you studied Lebesgue
integration at UMIST?'' When he heard that I had not, he said, ``In that
case, Mr. Tong, you are only half-educated. I suggest that you attend a
course on it in our inter-collegiate postgraduate programme.''

As there was no dedicated Lebesgue Integration in that year's
programme, I chose a course on Probability Theory (via Measure Theory).
The lecturer was none other than Professor Harry Reuter from Imperial
College, London. Much later I learned that he was famous for his
collaborative work with David Kendall on birth-and-death processes etc.
Again, he, the son of the Socialist Mayor of Magdeburg, came as a young
man to the UK to escape Hilter's Germany; he was looked after by the
Cambridge mathematical analyst, Professor Charles Burkill, and his
charitable wife, Greta Braun. Professor Reuter was such a wonderful
lecturer that he got me hooked. In fact, his course made me reconsider
my entire academic direction.

I decided that Probability would be far more fun and useful. The
decision to quit Algebra was not painful. One must always follow one's
passion. So, I~can honestly claim that I was facilitated by a famous
algebraist into Statistics. (In doing so, I~dropped from the 13{th}
generation of academic descendants of Sir Issac Newton to the
14{th}, according to the Mathematics Genealogy Project!) You
see, I have had experiences of discontinuous decisions more than once
in my life. Thresholds have been truly an integral part of my life in
more senses than one.

\begin{figure}[t]

\includegraphics{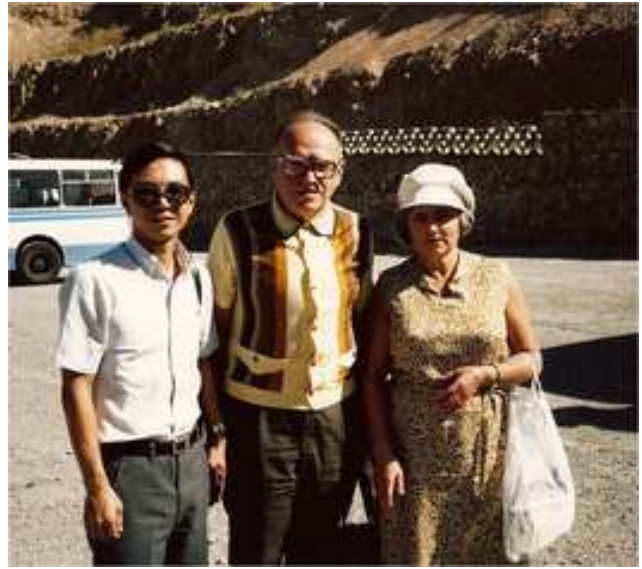}

  \caption{With Akiva Yaglom and his wife at the foot of the Tian Shan Mountain on the Tashkent side, 1986.}\label{fig2}
\end{figure}

As it turned out, I stayed at the Northern Polytechnic for just one
year. My teaching duty was not heavy and I had free time to read
around. I read several books on probability and stochastic processes.
For example, I~came across the delightful book on the theory of time
series by Akiva Yaglom, which kindled my interest in the subject. Many
years later, I was able to thank Akiva in person for his introduction.
I met him in 1986 at the First Bernoulli World Congress held at
Tashkent in the former USSR; we were both walking up the Heavenly
Mountain (or~Tianshan) from the Soviet side. We became instant friends.
Do you know that the theoretical underpinnings of the ARIMA model made
popular by George Box and Gwilym Jenkins were already laid rather fully
by him in 1955? I learned this fact from Peter Whittle's charming book
\textit{Prediction and Regulation}, first published in 1963, when he was
Professor of Statistics at Manchester. The book contains many gems and
has remained one of my favourites since my days at the Northern
Polytechnic. Another book that also captured my attention was the one
by Ulf Grenander and Murray Rosenblatt entitled \textit{Analysis of Stationary
Time Series} (1957). You know, in my day there were not too many books
on time series. One could probably count them on the fingers of one or
two hands.

\begin{figure*}

\includegraphics{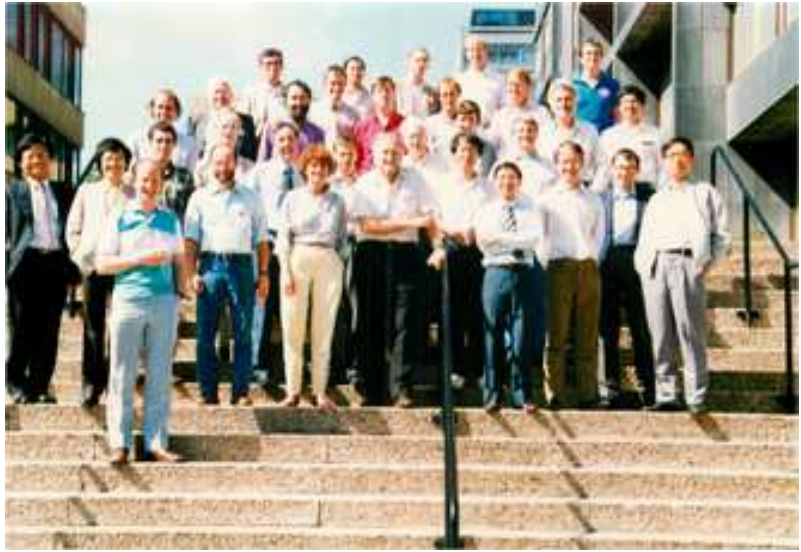}

  \caption{Edinburgh Workshop on Nonlinear Time Series Howell organised in 1989
(left to right ignoring row number: Wai-Keung Li, Ruey Tsay, Colin Sparrow, Russell Gerrard, John Lane, Murray Rosenblatt,  Gudmundur Gudmundsson, John Petruccelli, Tze-Liang Lai, Tony Lawrance, Peter Robinson, Dominic Guegan, T. K. Brown, Pham Dinh Tuan, Timo Terasvirta, Rodney Wolff, Clive Granger, Peter Fisk, David Cox, Martin Casdagli, Jonathan Tawn, Tohru Ozaki, Granville Tunnicliffe Wilson, Howell Tong, Dag Tjostheim, Ed McKenzie, Peter Lewis, Richard Smith, Neville Davies, David Jones, Kung-Sik Chan, Zhao-Guo
Chen).}\label{fig3}\vspace*{-3pt}
\end{figure*}

\begin{figure*}[b]

\includegraphics{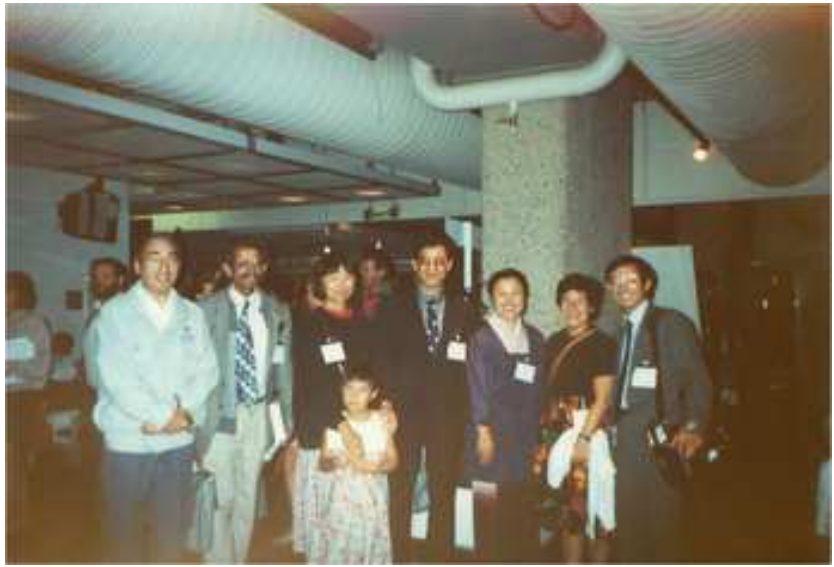}

  \caption{ISI meeting in Paris, 1989. Left to right: Maurice Priestley, Tata Subba Rao, Mary Tong, Anna Tong, Ritei Shibata,
Haruku Shibata, Nancy Priestley, Howell Tong.}\label{fig4}
\end{figure*}

\begin{figure*}

\includegraphics{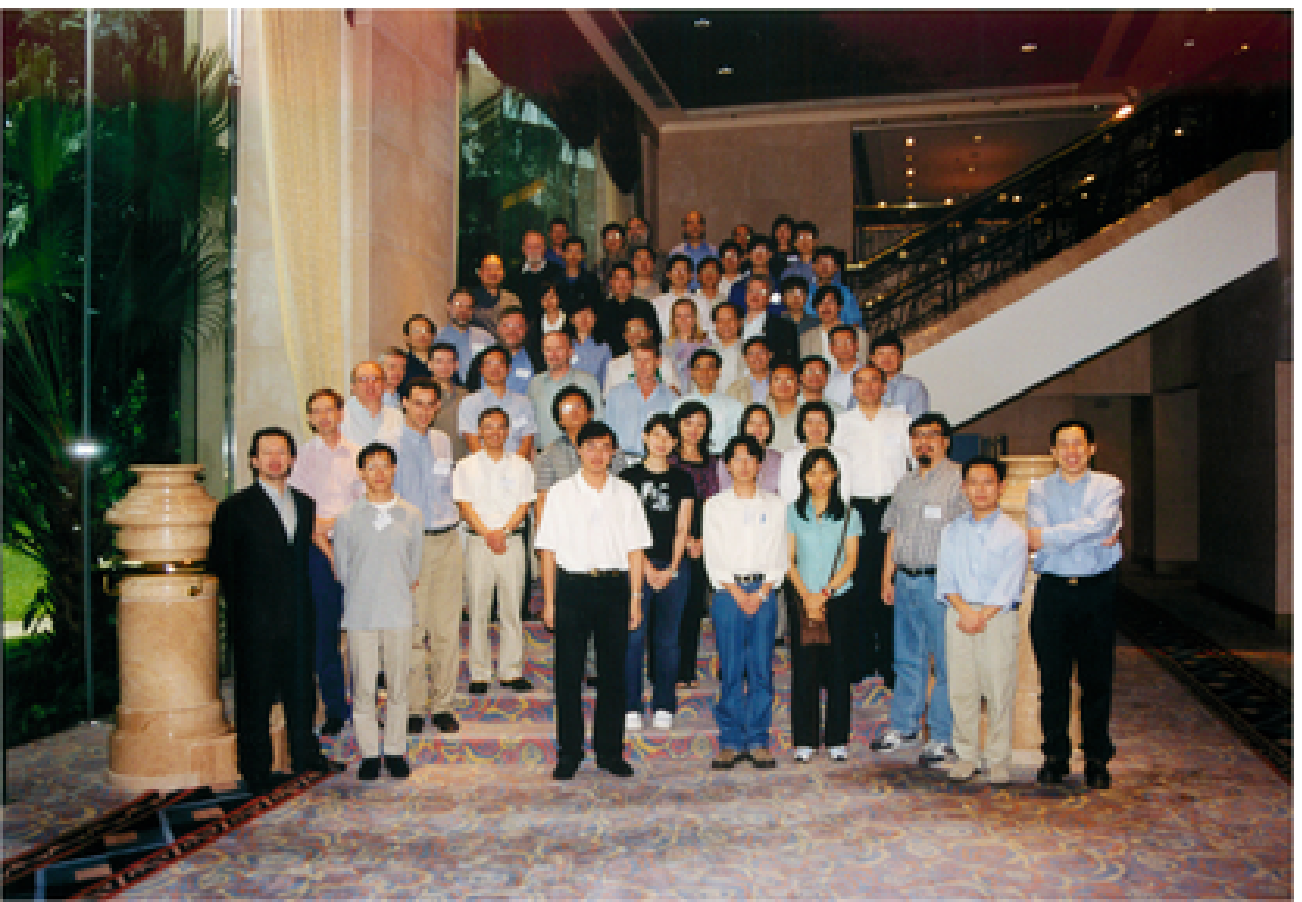}

\caption{International Conference on Financial Statistics, Hong Kong, 1999.}\label{fig5}
\end{figure*}

At the Northern Polytechnic, there was then a small study group on
forecasting led by Dr. Warren Gilchrist, who later moved to head the
Statistics Department at the Sheffield Polytechnic, now called the
Sheffield Hallam University. I went along mainly to listen. Then one
day I was asked if I would like to speak to the group on a paper of my
choice. I~happened to be studying Jim Durbin's \textit{Biometrika} paper on the
fitting of a moving average model via a long autoregression. I remember
showing the group all my calculations, which helped me understand the
paper and survive my first seminar. Little could I foresee at the time
that my path would cross Durbin's several times later in my life. When
Priestley's name was mentioned at one of the meetings of the study
group, I looked up some of his papers, after which I knew that I would
have to return to UMIST!

You see, Maurice came to UMIST just when I was starting my final
undergraduate year; he lectured to us on mathematical statistics and
stochastic processes. We at UMIST had excellent exposure to Statistics
through Peter Wallington and Maurice Priestley. The former worked on
queuing theory under Dennis Lindley. The only trouble was that they
made the subject LOOK so easy that two of the more academically
inclined students, including myself, opted for something more abstract
like Algebra!\looseness=1

To cut a long story short, Maurice welcomed me back. In fact, thanks to
an oversight on the part of the head of department (Maurice was not the
Ho.D.), I~was appointed as a demonstrator to compensate for the SERC
postgraduate studentship that the Ho.D. forgot to apply on my behalf. The
upshot was that I started my university teaching career as a
postgraduate student and joined the university pension scheme at quite
a young age. This turned out to be very beneficial many years later
when my university pension (based on defined benefits) was calculated.

 \textbf{QY:} What made you shift from frequency-domain to
time-domain in your research in time series analysis?

 \textbf{HT:} As we all know, the history of time series
analysis switches to and fro between the time domain and the frequency
domain. I started my research from the frequency-domain end. I stayed
with it for a few years. Then in 1973, Maurice, Subba Rao and I got a
research grant, with which Professor Hirotugu Akaike of Japan was
invited to visit us for six months. Hiro's visit marked the beginning
of the end of my frequency-domain research. Let me elaborate.

The first phase of Hiro's time series research had been almost
exclusively frequency-domain. He was in fact an international figure in
the area. Then he started his collaborative research in designing a
feedback controller for a cement kiln. To his dismay, he discovered
that in the presence of feedback, the frequency-domain approach was
inadequate due to a serious bias problem associated with the estimation
of the frequency-response function. His findings were recorded in the
\textit{Proceedings of Spectral Analysis of Time Series} edited by Bernard
Harris in 1967. This impressive piece of work led to the invitation
from UMIST.

His visit gave me ample opportunities to learn from his experiences. He
was working on his fundamental state--space work at the time, which
culminated in identifying a state as a basis vector of the predictor
space of a second-order stationary multivariate time series. His vast
knowledge impressed me deeply, so I decided to visit his institute in
Tokyo, Japan. He was very supportive of my wish. In the event, I was
awarded a Royal Society Japan Fellowship without any trouble. I guess
that I could well have been the only applicant, as the fashion of the
day in the UK was to go westwards. The six months I spent at Hiro's
institute completed my (inverse) Fourier transform and I returned to
the UK as a predominantly time-domain person. I have already related
the transformation process in my obituary of Professor Akaike published
by both the Royal Statistical Society and the Institute of Mathematical
Statistics. Therefore, I shall not repeat the account here, except to
say that his personal mini-library played a vital role.

 \textbf{KSC:} Your earlier works in time series analysis
were all linear. What made you decide to switch to nonlinearity?

 \textbf{HT:} Again it had to do with an RSS discussion
meeting. On 18th May, 1977, I read a very short paper to the RSS, as
one of three discussion papers. At the meeting, two features were
highlighted, namely, time-irreversibility and limit cycles. I can
remember the challenging problem posed by Dr. Granville Tunnicliffe
Wilson: ``Would we not prefer a model which in the absence of such (he
meant random) disturbances would exhibit stable periodic deterministic
behavior---a limit cycle?'' I decided to take up the challenge.

Coincidentally, around the same time, the Swedish control engineer,
Professor K. Astr\"{o}m, gave a seminar at UMIST. He described a
bilinear control system, in which the output is not just a simple
linear function of past (control) input and past output but also their
cross products. For time series analysts, an obvious way to imitate
this framework is by replacing the control input by a stochastic noise.
(Of~course, in doing so we are replacing a manipulated variable by an
unobservable one!) I played around with this idea for a bit and even
published something on it.

\begin{figure}

\includegraphics{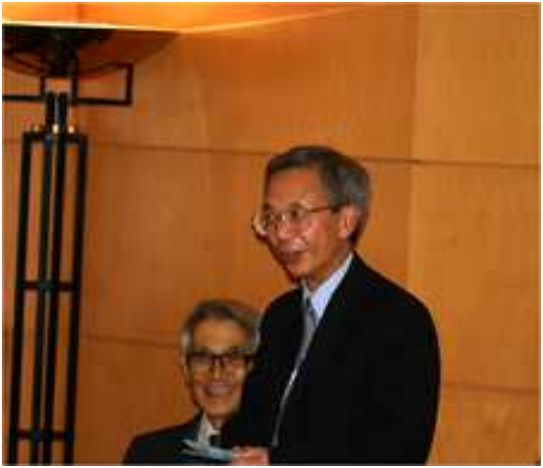}

\caption{Hirotugu Akaike enjoying Howell's after-dinner speech at a conference honoring Akaike, Yokohama, 2003.}\label{fig6}
\end{figure}

However, very quickly I convinced myself that was probably not the best
way to address Granville's challenge: if I switch off the driving
noise, the system would grind to a halt! One day, as I was mowing my
lawn, strip by strip, it dawned on me that a piecewise linear model
would be a good candidate. The rest is history, which you know I have
recounted in the article ``Birth of the threshold time series model'' in
\textit{Statistica Sinica} (2007).

Actually, the earliest mention of the idea can be traced to my
contribution to the discussion of Tony Lawrance and N. T. Kottegoda on
modelling of riverflow time series in 1976. There was an interesting
follow-up. At the time, it seems that my friend Tony could not see any
relevance of the threshold idea to riverflow time series modelling. I
am sure this was my fault. So, understandably he complained that I and
one other contributor were ``following a tradition of the Society in
taking the opportunity to publicize their forthcoming works---at the
expense of other authors' reprint charges.'' I hope that subsequent
applications of the threshold model in riverflow time series modelling
and linking of the Lawrance--Lewis's exponential autoregressive model to
the threshold model have convinced him that the additional reprint
charges were perhaps not unjustified.

 \textbf{KSC:} Can you tell us more about the development of
the threshold models, including their impact on ecology, economics and
finance and other areas?

\begin{figure}

\includegraphics{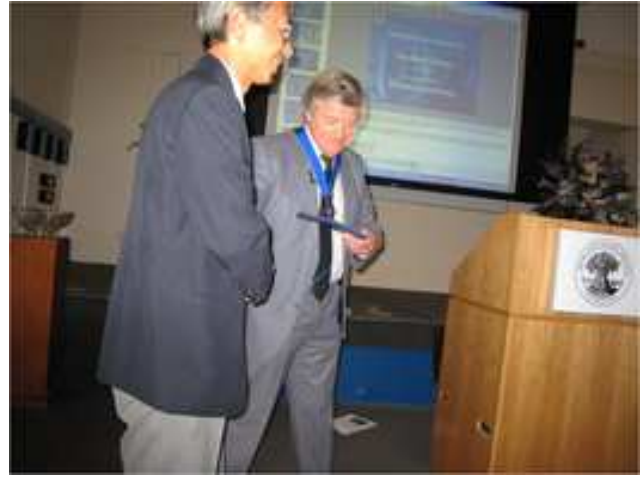}

\caption{Howell receiving the 2007 Guy Medal in Silver from President Tim Holt.}\label{fig7}
\end{figure}

 \textbf{HT:} I have given a fairly detailed overview in my
article ``Threshold models in time series analysis---30 years on'' in
\textit{Statistics and Its Interface} (2011). I~sincerely hope that the model
will continue to enjoy its popularity with users from diverse
disciplines. It makes me a very happy man when I see applications of
the model in econometrics, economics, finance, ecology, epidemiology,
psychology, hydrology and many others. Frankly, some of the application
areas are beyond my wildest dream. For example, just the other day my
attention was drawn to cover song detection and bipolar disorder via
the threshold model.

\begin{figure}

\includegraphics{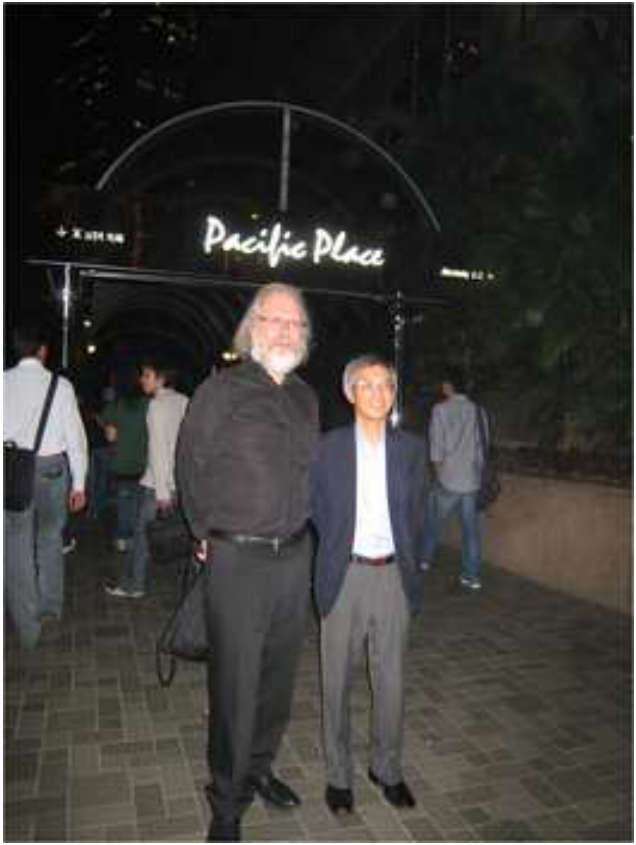}

\caption{Nils Christian Stenseth and Howell, in Hong Kong, 2008.}\label{fig8}
\end{figure}

It would be wonderful if somebody could put all the most successful
applications in book form. Hint, hint\ldots

Now the basic idea of the threshold model is very simple: divide the
state space into regimes and rule each with a simple linear model. It
has a nonparametric flavor within a parametric framework. Of course, if
we divide the state space arbitrarily finely, as in a spline approach,
we gain generality at the expense of loss of parsimony or
interpretability. Successful applications of the threshold model have
shown that, in many real applications, two or three regimes will often
suffice. Especially encouraging is the fact that quite often the
regimes are interpretable. In mathematics, the idea of piecewise
linearization is, of course, very old. In oscillations theory, the
former Soviet mathematicians, Andronov and Khaikin, had introduced and
studied (nearly) exhaustively piecewise linear differential equations
in the 1930s. In statistics, we had two-phase linear regressions and
Tukey's regressogram a long time ago, but it seems that they had made
no or little impact on time series modelling, till the launching of the
threshold autoregressive model and more generally the threshold
principle. I must say that from the standpoint of stochastic dynamical
systems, the \mbox{incorporation} of time in a regression framework is a
paradigm-shifting step because without time there is no dynamics. This
is why I hail Yule's invention of the autoregressive model as one of
the greatest revolutions in statistical modelling because it ushered in
the era of dynamic (as against static) modelling. I~find it unfortunate
that some recent textbooks have blurred the distinction between a
dynamic model and a static model.

Bruce Hansen (\citeyear{Han11}) has given an extensive review of the impact of the
threshold model in econometrics and economics. Without any doubt, it is
in econometrics/economics that the threshold model has made its
greatest impact. More recently, the influence seems to be spilling into
the field of finance including actuarial science.

\begin{figure}

\includegraphics{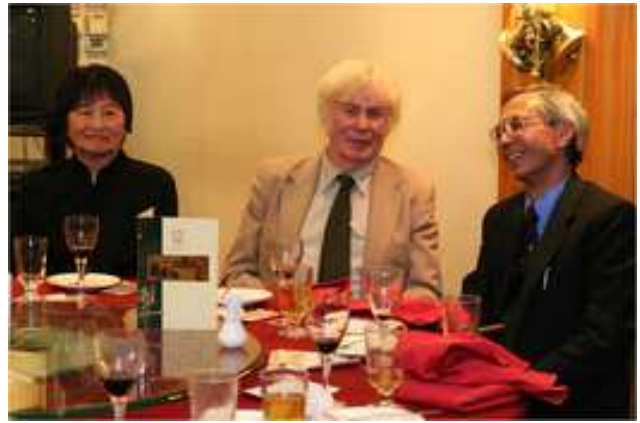}

\caption{Mary, Peter Whittle and Howell, in Hong Kong, 2009.}\label{fig9}
\end{figure}

Another significant area of application is ecology. Of course, you,
Kung-Sik, have done some marvellous joint work with our dear friend,
Nils Christian Stenseth of Norway. You have covered so much of the
animal kingdom: mink, lynx, rodent, lemming and so on. Your more recent
work with your former doctoral student, Noelle Samia, and Nils
Christian's team on plague epidemics using data from Kazakhstan is
truly wonderful. As your papers have shown yet again, often it is
through real applications that real progress on the implementation of
what I have called the Threshold Principle can be made. You have
implemented the principle for count data. I don't want to embarrass
you, Kung-Sik, but I must say that the implementation is a truly
remarkable contribution.

\begin{figure*}

\includegraphics{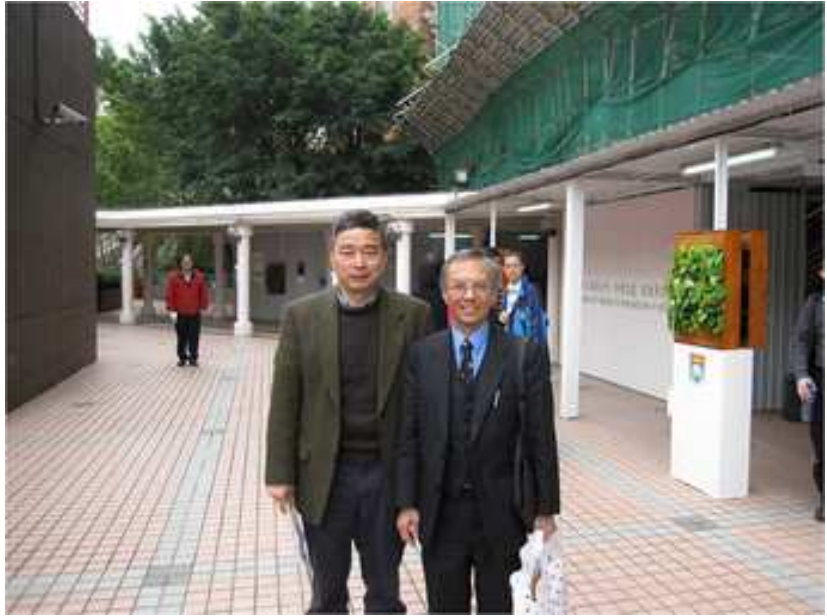}

\caption{Qiwei Yao and Howell in Hong Kong, 2009, with  Wai-Keung Li and Mike So in the background.}\label{fig10}
\end{figure*}

Of course, regimes can be delineated either sharply or smoothly. Coming
from Hong Kong, I am rather happy with a sharp border! Well, the
self-exciting threshold autoregressive (SETAR) model uses a sharp
delineation. However, some people are less receptive to sharp
delineations. In this case, we can consider a softer delineation, for
example, a smooth (perhaps ``soft'' is a better word) threshold
autoregressive model. You, Kung-Sik, and I have actually developed
quite a comprehensive methodology and we have even given it the acronym
of STAR model.

The idea has apparently attracted considerable attention in the
econometrics literature, under the same acronym. I could perhaps make
one or two remarks here. For simplicity, let us consider a
one-threshold model. If the estimated threshold is in the vicinity of
small probability, for example, near the tail or an anti-mode of the
marginal distribution, then it tells us that there is probably
insufficient information in the data on the functional form of the
model there. In that case, whether we use an indicator function as in
the SETAR model or a more sophisticated smooth function as in the STAR
model is of secondary importance. After all, all models are wrong. When
choosing between a SETAR model and a STAR model, a more relevant
question is which one is more useful and interpretable.

More recently, you, Kung-Sik, Shiqing Ling, Dong Li and I have shown
systematically how the threshold approach can provide powerful tools to
model conditional heteroscedasticity in finance, environment, ecology
and others. We have exploited the mixture of distributions in the
driving noise of the threshold approach.

So far I have focused my answer on a univariate time series. Although
there are generalizations of the threshold model to multivariate time
series, I think much work remains to be done. One key question is the
delineation of regimes for a $p$-dimensional state space. The
topography can be quite vast. Too vast perhaps? My gut feelings are
that it is still possible to construct an efficient search algorithm.

Besides the question of sharp and smooth delineation, there is also the
one to do with observable or hidden threshold variables. I must tell
you that I wasted an excellent research problem of Markov-chain driven
TAR model in 1983 by assigning it to the wrong student; I should have
passed it to you, Kung-Sik, and you would have cracked it in three
months. The idea was there in
the paper I read to the RSS in 1980 (page 285, line 12 from below).

\begin{figure*}[b]

\includegraphics{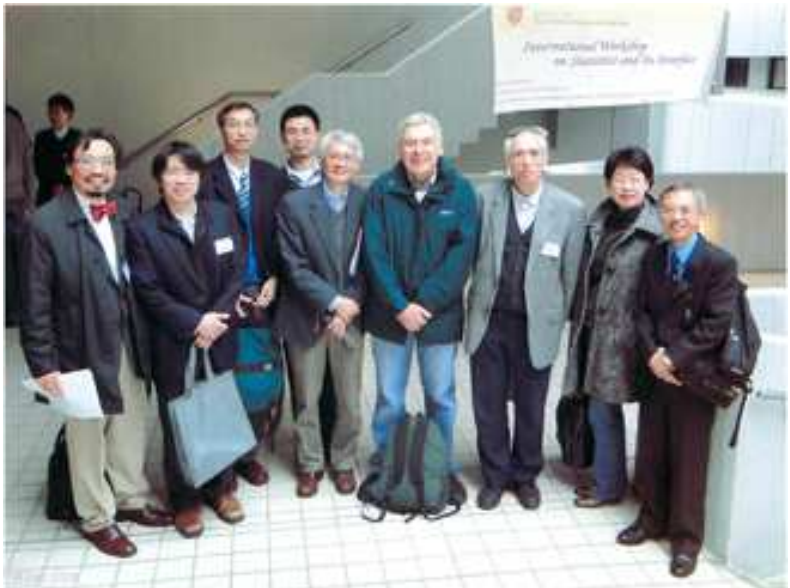}

\caption{P. S. Wong, C. K. Ing, N. H. Chan, W. Wu, K. L. Tsui, Peter Hall, T. L. Lai, R. Liu and Howell, at the Chinese University of Hong Kong, in 2009.}\label{fig11}
\end{figure*}

Sometimes, we can even consider partially observable and partially
hidden threshold variables. I have given a discussion in my 2011
recount in \textit{Statistics \& Its Interface}.

 \textbf{KSC:} On looking back, the threshold idea is very
natural. Nowadays the idea is applied in many areas, for example,
ecology, economics and so on. And the TAR models are often featured
very substantially in elementary text-books, for example, Walter
Ender's \emph{Applied Econometric Time Series Analysis} and Cryer and
Chan's \emph{Time Series Analysis: With Applications in R}. Yet, the
idea seems to have taken quite some time before it was universally
accepted. Don't you think that this is a little odd?

 \textbf{HT:} Well, it was probably my fault as much as yours
for not being good salesmen! More seriously, as I have hinted at
earlier, the history of statistics is full of cases of belated
recognition as well as premature enthusiasm. Of course, there are also
cases of instant recognition that have withstood the test of time. Like
many other professions, value judgments by statisticians can sometimes
be more subjective than scientific. I prefer to let TIME be the judge.
I~can remember Hiro Akaike saying to me many years ago (perhaps it was
in the 1970s), ``I reckon that AIC could probably survive 30 years.'' You
see, even he had made the wrong prediction about his own baby!

 \textbf{QY:} You have also had keen interest in chaos. How
does chaos fit in with statistics in general and time series in particular?

 \textbf{HT:} The primary object of study in Statistics is
chance or, equivalently, randomness. The traditional view in statistics
seems to place randomness at one end and determinism at the other. And
it would be heresy to mix the two. In fact, every statistician carries
with him $\varepsilon$'s everywhere, as if he owes his entire existence
to them. If you ask him where his $\varepsilon$'s come from, he would
give you a long list of sources, which is usually all right as far as
it goes, except for the likely absence of one very significant
ingredient. Let me digress first.

Suppose I toss a coin in this room. I hope you will agree that it is a
reasonably close system free from external disturbances. Now, I can
write down the precise equations of motion of the coin by appealing to
Newtonian mechanics. But I also know I cannot predict its outcome with
certainty, if I give it a good throw. Why? Where is the source of
randomness? As long ago as the beginning of the 20{th}
Century, H.~Poincar\'{e} already included sensitivity to initial
conditions as a significant source of randomness. So, even the most
basic generator of randomness used by a statistician is a deterministic
system; its randomness is due to what is called chaos by the
dynamicists. Thus, what excuses can statisticians have to ignore chaos?
Rather than burying our heads in the sand, I suggest that it is more
constructive for us statisticians to learn more about chaos and make
our contributions. Another interesting example is to do with point
processes. Within the setup discussed in David Cox and Walter Smith
(\citeyear{CoxSmi54}), we can identify a connection between point processes and chaos
via the circle map: $x_n = x_{n-1} + \Theta$, $x_0 = 0$ ($n=1,2,\ldots$),
where we observe $y_n = x_n  \operatorname{mod} 1$. Note that for irrational
$\Theta$, $y$ is uniformly distributed on $[0,1)$. I referred to this
connection in my reply to David in my 1995 discussion paper in the
\textit{Scandinavian Journal of Statistics}.

You asked about time series. It turns out that many nonlinear time
series models in statistics do generate chaos when we switch off the
driving noise. That is what makes them so endearing! In a sense, there
is the inherent randomness due to chaos of the underlying deterministic
system (I have called it the skeleton elsewhere), as well as the other
randomness due to the random driving force, perhaps reflecting the fact
that we are dealing with a complex system with multiple sources of
randomness, some, but usually not all, of which can be explained with
some degree of precision.

\begin{figure*}

\includegraphics{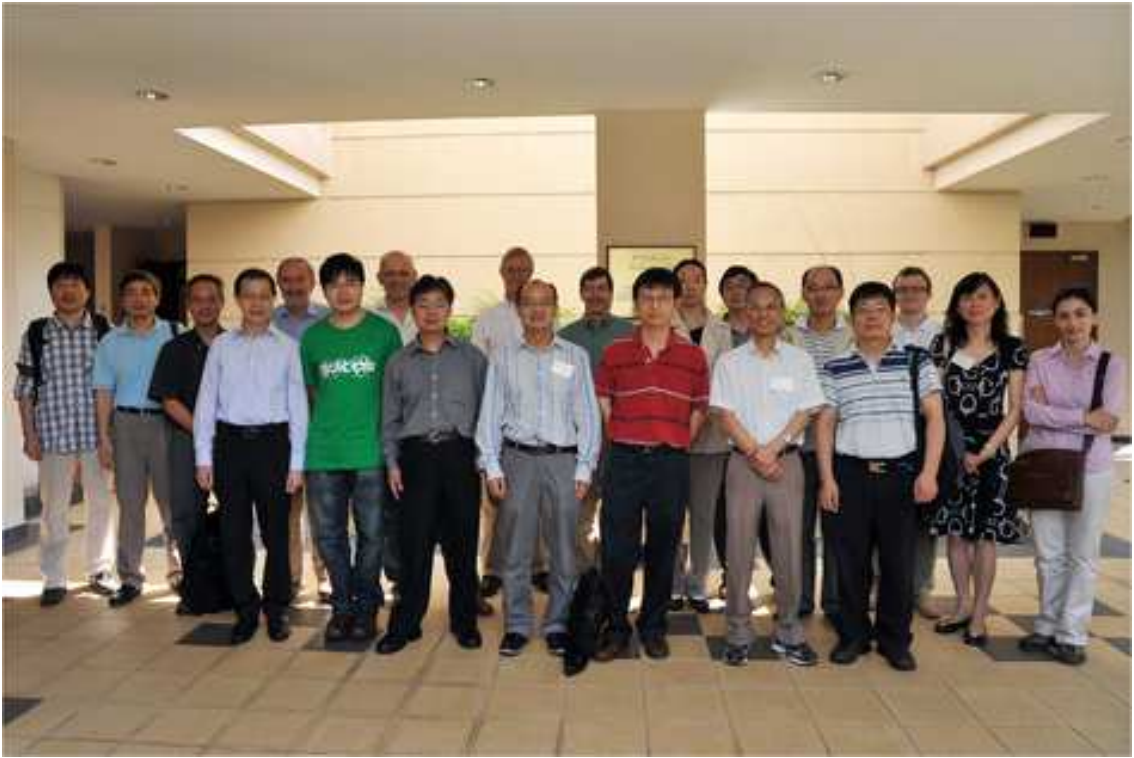}

\caption{Howell with colleagues at the Nonlinear Time Series Workshop in National Singapore University, 2011; from left to right and ignoring rows: Dong Li, Qiwei Yao, Kung-Sik Chan, Mike So, Peter Brockwell, Ken Siu, Rainer Dahlhaus, Zudi Lu, Marc Hallin,  Cheng Xiang, Richard Davis, Yingcun Xia, Ying Chen, Rong Chen, Howell Tong, Myung Seo, Shiqing Ling, Simone Giannerini, Cathy Chen, Azam Pirmoradian.}\label{fig12}
\end{figure*}

If we accept the above argument, then a natural question is how to
define initial-value sensitivity of a \emph{stochastic} dynamical
system. Of course, Qiwei, you know the answer very well, as we have
written about the topic. It turns out that the conventional approach
adopted by the deterministic dynamicists is inadequate, as it ignores
the diffusion due to the existence of multiple sources of randomness.
Instead of looking at the movement of state $x$ from one time instant
to the next as they do in deterministic dynamics, we now look at the
movement of one distribution $F(x)$ from one time instant to the next.
Since the focus is now on the distribution, we have to generalize the
way we measure the sensitivity of the movement to initial values (i.e.,
initial distributions). We introduced a stochastic counterpart of the
Lyapunov exponent. This experience shows the benefit of having
statisticians involved in the study of deterministic chaos.

 \textbf{KSC:} You interacted with people outside statistics.
How did that come about?\vadjust{\goodbreak}

 \textbf{HT:} Mostly by chance and more importantly by taking
advantage of it. It is important to enjoy listening and have a sense of
curiosity. For example, I~collaborated with Dr. Gudmundsson of Iceland
because I remembered that he was working on geophysical problems when
he was a post-doctoral research fellow at UMIST. I met him there when I
was a research student, and I listened to him and remembered what he
had told me. So, many years later, I contacted him when I was
interested in riverflow time series. Another example is Professor Nils
Christian Stenseth. I~met him via his doctoral student Ottar Bj\o
rnstad, who contacted me and invited me to visit his department. I went
to Oslo, listened to him and his colleagues and found the team there
ideally placed for collaborative research. Nowadays, the internet is
wonderfully convenient. Sometimes, I~have not even ever met my
co-authors in person.

 \textbf{QY:} Besides time series analysis, you have also
worked in other areas of statistics, for example, Markov chain
modelling, reliability, dimension reduction. What motivated you?

 \textbf{HT:} They were mostly my part-time activities for a
bit of fun, except for dimension reduction, which was serious business.
By about the mid-1990s, I~knew I had to get into nonparametrics and
semi-parametrics. But they were developing very rapidly. It was not
easy for me to keep up, especially at a time when I was heavily
involved with administration. Luckily, Bing Cheng and you, Qiwei,
arrived in Canterbury, UK. I have learned so much from you. Thank you
very much!
As for dimension reduction, there is an interesting story behind it. As
you know, the area actually laid outside my normal expertise in the
1990s. I was starting my sabbatical leave at the University of Hong
Kong from the University of Kent, UK, initially for three years---I was
lucky. I~knew that Dr. Lixing Zhu of the department (now chair
professor at Baptist University, Hong Kong) was an expert in
semi-parametrics. So, I discussed dimension reduction with him. I was
not impressed with the need in the literature to under-smooth the
estimator of the nonparametric function. It might also be then or
perhaps a little later when I questioned the efficacy of techniques
such as the sliced inverse regression estimation for time series
because time-irreversibility is the rule in real time series. Lixing
shared my concerns but was himself very busy with other research
problems, so he mentioned the problem to one of Professor Wai-Keung
Li's new research students, Yingcun Xia. Yingcun was an exceptionally
bright student. To cut a long story short, his doctoral thesis formed
the basis of a joint discussion paper on MAVE which I, on behalf of the
four authors, read to the RSS in 2002. The trick was to estimate both
the nonparametric part and the parametric part jointly. In this way
under-smoothing is rendered unnecessary.

\begin{figure}

\includegraphics{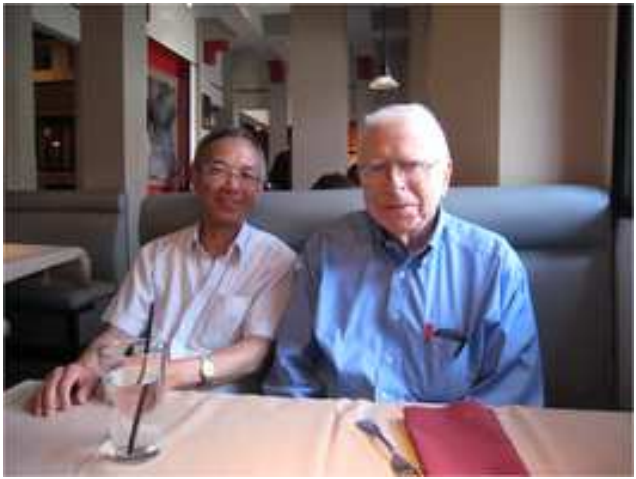}

\caption{Howell  and Murray Rosenblatt, after the former received the Distinguished Achievement Award from the International Chinese Statistical Association at the Joint Statistical Meeting at San Diego, USA, in July 2012.}\label{fig13}
\end{figure}

 \textbf{KSC:}
We all know that you have held senior administrative positions in five
universities across two continents. Can you share your experience with
us please? Perhaps you could begin with the Chinese University of Hong Kong.

 \textbf{HT:}
After working at UMIST for 14 years, I~thought it was high time for me
to return to my birth place, Hong Kong. There was a newly created
Department of Statistics at CUHK around 1981 and a new chair of
statistics was advertised, to which I applied successfully. The new
department in 1982 consisted of 5 faculty members including myself, one
senior lecturer and three lecturers. (CUHK followed the British system
at that time.) There were also one assistant computer officer (that was
you Kung-Sik),
\begin{figure*}

\includegraphics{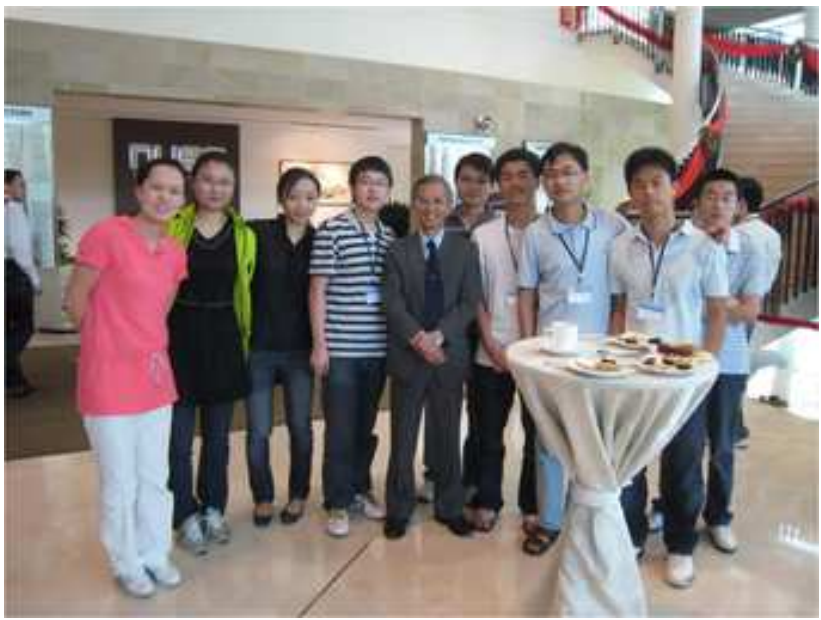}

\caption{Howell with a group of post-graduate students at
National University of Singapore, 2012.}\label{fig14}
\end{figure*}
one secretary and one messenger boy. Although I was the
founding chair professor, actually I did not appoint them; all the
faculty members were transferred from the Department of Mathematics and
all the lecturers were formerly students of the senior lecturer.
Fortunately we got on very well indeed. Staff and graduate students had
regular dim-sum lunches at a local restaurant. We shared the cost, the
seniors paying more, of course---a~workable socialist system! The
biggest challenge was actually curriculum design. We decided that our
first year undergraduates should receive good groundings in the guiding
principles of our subject rather than routine mathematical
manipulations. I was voted to be the guinea pig. It was fun and I
learnt a lot myself! Professor George Tiao was our external examiner
(another British practice) and he was most helpful and supportive. He
made plenty of constructive suggestions and gave us every
encouragement. He has been maintaining excellent relationship with CUHK
and many other tertiary institutions in Hong Kong ever since.

 \textbf{KSC:}
What made you decide to leave CUHK in 1986?

 \textbf{HT:} My decision to leave CUHK had nothing to do
with local politics of the time. I was quite happy at CUHK and my
vice-chancellor (equivalent to a university president in the US) was
very happy too with the development of my department and the department
has remained in very good shape to this day. In fact, it all happened
quite by chance when I was visiting Professor David Cox's department at
Imperial College, London. One day, David told me that a chair was to be
advertised by the University of Kent at Canterbury, UK. He suggested
that I could have a go if I was interested in returning to the UK.
Well, I do not know to this day why UKC decided to appoint me instead
of any one of three other very strong candidates.
As it turned out, the biggest challenge was how to manage a not so
united mathematics department, consisting of pure mathematicians,
applied mathematicians and statisticians. There were three sections,
three budget holders and all in one department. A bit crazy! A year or
two after my arrival, the vice-chancellor appointed me as the director
of my department (directorship was by appointment then). When I became
aware of the wish of the university to build up statistics and
actuarial science by running down (pure and applied) mathematics, I
reminded the vice-chancellor first the history of Thomas Becket and
then my plan. As the director of my department, I could not possibly
run down two sections to fatten up the third, especially when the
latter was associated with me. However, I could build up statistics
without harming mathematics by (i) taking advantage of the donation
secured by my predecessor from the Black Horse financial group to build
a solid base for actuarial science; (ii) taking over a major portion of
the management science group which was being or about to be
re-organised; (iii) consolidating statistical consulting activities and
service provision to Pfizer, whose UK base was nearby. By the time I
stepped down as director in 1993, the statistics group (including
actuarial science and the consulting arm) grew to more than 30
full-time staff working under one roof, possibly the biggest in the UK
then. Our research rating also went up from 2 when I joined to 4 when I
stepped down.

 \textbf{QY:} Yes, I can remember those exciting days when I
joined you in 1990. Then you went to Hong Kong in 1997. Can you take us
through that period please?

 \textbf{HT:} Again it was purely by chance that I went to
Hong Kong, this time to the University of Hong Kong. You see, HKU had a
new and very enterprising vice-chancellor, Professor Patrick Cheng. He
was working very hard to turn HKU from a sleepy teaching-oriented
university created in the colonial days to a research-vibrant modern
university. He was investing huge resources in attracting people from
all round the world to HKU by creating positions such as distinguished
visiting professorships. A long-time fellow time series analyst, Dr.
(now Chair Professor) Wai-Keung Li, seized the opportunity and was
instrumental in getting me appointed. I arrived in HKU in 1997 on a
3-year sabbatical leave (without pay, of course) from UKC. At that
time, UKC also had a new vice-chancellor, Professor Robin Sibson. It
was he who granted me the leave.

 \textbf{QY:} You were a visitor and yet you became the
founding dean of their graduate school. How did that come about?

 \textbf{HT:}
Well, it was all due to my big mouth as usual. My perpetual problem!
After my arrival at HKU, one morning Wai-Keung (who was HoD) said to
me, ``Howell, as you are a chair professor, I'd suggest that you attend
our senate meeting this afternoon if you can spare the time. You see, I
cannot go because I have some departmental matters to attend to.
Anyway, it might be fun for you to see how we operate at HKU.'' It
turned out that the controversial item on the agenda was the
establishment of a graduate school at HKU. The debate was getting
really heated. It did not take me long to realize that many of those
who opposed setting up a graduate school were professors who came from
Britain ten or twenty years previously during the colonial days. You
can tell from their accents! I could see that the vice-chancellor and
his team were getting nowhere. At this point, I thought I had to say
something. So, I said, ``As somebody who has just arrived from Britain,
I would like to inform senate members, especially those who left that
country many years ago, that the concept of a graduate school, no doubt
an American concept, is being adopted by a rapidly increasing number of
universities in Britain. I feel that this is an irreversible trend
world-wide.'' After that, the debate subsided and the motion was
carried. The following morning, the vice-chancellor rang me up. After
thanking me for my intervention, he invited me to be the founding dean.
The rest is history. My wife joked with me afterwards, saying ``I
thought you wanted to escape to Hong Kong in order to have peace and
quiet. See what you have done. Serves you right with your big mouth!''
Well setting up a graduate school at HKU was challenging, because my
first job was to persuade nine faculties to relinquish their power to
the graduate school, abide by some common rules and regulations and to
accept supervision by the Graduate School. I had two associate deans
(Professors John Malpas and Anthony Yeh) and one senior administrator
(Mrs. Yvonne Koo) from the registry to assist me---we called ourselves
the gang of four. We literally set down all the rules and regulations,
down to the way we handled reference letters. We always sent a
thank-you letter to each referee enclosing a copy of his/her reference
letter. This is a good way to uncover monkey business. In just a few
years, we succeeded in improving our thesis completion rate (after
constant monitoring of progress) and employability of our graduate
students (we ran a small number of compulsory language-enhancing and
skill-empowering courses).

 \textbf{KSC:} And you also became a pro-vice-chancellor
(equivalent to a vice-president in the US system)!

 \textbf{HT:} Yes, I did serve as PVC to three VCs at HKU. My
portfolio changed from one VC to the next and it included, at different
times, research, administration and development. The names did not mean
much because the dividing line was not sharp. My research portfolio did
mean that I was in charge of the university's all important submission
of research output to the Hong Kong University and Polytechnic Grants
Committee, who decides our budget. The work was tedious but it had to
be done methodically and colleagues had to be handled delicately and
with compassion. I remember visiting a number of departments and
chatting to all the 60 or so heads of departments.

 \textbf{KSC:} You have collaborated with many people, mostly
younger than you, in research. Can you share your experience with us?

 \textbf{HT:} I have always enjoyed young companies. They are
without baggage, full of vitality and can think the unthinkable. My
experience suggests to me that it is far easier sharing crazy ideas
with the young than with the old. The old tends to react almost
immediately by saying, ``They are wrong'' or ``They are trivial.'' But the
young would say, ``Oh, that is interesting. Let's see!'' I also think
that it is the duty of every statistician to work, from time to time,
with somebody younger than himself, for otherwise there is no hope for
the profession.

 \textbf{QY:} Now that you have retired from the London
School of Economics, how do you occupy your time?

 \textbf{HT:} Now that I have retired from the chair from
which Professor Jim Durbin also retired, it seems that I am as busy as
ever. The freedom from administration has given me more time to think
(hopefully deeper), travel and try other things. (I did enjoy
administration when I had to do it. You see, I saw no point in
complaining and making myself miserable.) Now, with my wife suddenly
becoming a qualified keep-fit instructor in her retirement, I have been
persuaded to exercise more regularly than I used to. I~also try to keep
up with the statistical literature and continue doing some research. I
am not displeased with some of the recent results I shared with young
colleagues. As a matter of fact, Yingcun and I published a discussion
paper in \textit{Statistical Science} in 2011. We argue that, for dependent
data, the MLE and its equivalents are not necessarily the most
efficacious when we know that the model is wrong. For example, for a
wrong time series model, conventional methods still typically rely on
functionals of the one-step-ahead predictors. We have challenged them.
More recently, Kung-Sik, Shiqing Ling, Dong Li and myself have just had
our paper on conditionally heteroscedastic AR models with thresholds
accepted by \textit{Statistica Sinica}, to do with the threshold approach.

I have joined the University for the 3{rd} Age, through which
I have participated in activities that I have never imagined I could
do. For example, I~enjoyed the course on book-binding. In fact, I~have
turned my copy of Peter Whittle's charming little book \textit{Prediction and
Regulation} from a poorly produced paperback version into an acceptable
hardback. Do you know that Peter is also a bookbinder? I discovered
this fact when I showed him the finished product. Moreover, I am now
able to indulge myself more in History, Literature and Philosophy. One
regret is that I am not trilingual or better. I~would love to be able
to enjoy, for example, \textit{War and Peace} in Russian. So much is often
lost in translation. Just compare Witter Bynner's translation (possibly
the best available):

\begin{quote}
``\ldots Though I have for my body no wings like those of the
bright-coloured phoenix,

Yet I feel the harmonious heart-beat of the Sacred Unicorn\ldots''
\end{quote}
with the famous original verse of Li Shangyin (ca. 813--858).

I have digressed!

To me, retirement is one LONG (I hope) sabbatical leave that has opened
doors into many fascinating avenues. I recommend it!



%

\end{document}